\documentclass[twocolumn]{aastex701}
\usepackage{color}
\usepackage{amsmath}

\newcommand{\eg}{e.g.,\ }
\newcommand{\etal}{et~al.\ }

\newcommand{\magsec}{mag arcsec$^{-2}$}
\newcommand{\Msun}{$M_\odot$}

\newcommand{\hi}{\ion{H}{1}}

\begin{document}

\title{Stellar Populations in the Extreme Outer Halo of the Spiral Galaxy M96}
\shorttitle{Stellar Populations in the Halo of M96}
\shortauthors{Mihos \etal}

\author[0000-0002-7089-8616,gname='Chris',sname='Mihos']{J. Christopher Mihos}
\affiliation{Department of Astronomy, Case Western Reserve University, Cleveland OH 44106, USA}
\email{mihos@case.edu}

\author[0000-0001-9427-3373]{Patrick R. Durrell}
\affiliation{Department of Physics, Astronomy, Geology, and Environmental Sciences, \\
Youngstown State University, Youngstown, OH 44555 USA}
\email{prdurrell@ysu.edu}

\author[0000-0002-8543-6406]{Brian Malkan}
\affiliation{Department of Astronomy, Case Western Reserve University, Cleveland OH 44106, USA}
\email{brmalkan@gmail.com}

\author[0000-0003-4859-3290]{Aaron E. Watkins}
\affiliation{Centre for Astrophysics Research, University of Hertfordshire, College Lane, Hatfield AL10 9AB, UK}
\email{a.emery.watkins@gmail.com}

\begin{abstract}

We use deep {\sl Hubble Space Telescope} imaging to study stellar populations in the outer halo of the spiral galaxy M96, located in the dynamically active Leo I galaxy group. Our imaging targets two fields at a projected distance of 50 kpc from the galaxy's center, with a 50\% photometric completeness limit of F814W = 28.0, nearly two magnitudes below the tip of the red giant branch. In both fields we find a clear detection of red giant stars in M96's halo, with a space density that corresponds to an equivalent broadband surface brightness of $\mu_V \approx 31.7$ \magsec. We find little evidence for any difference in the spatial density or color of the RGB stars in the two fields. Using isochrone matching we derive a median metallicity for the red giants of $[M/H] = -1.36$ with an interquartile spread of $\pm0.75$ dex. Adopting a power-law radial density profile, we also derive a total halo mass of $M_h = 7.8^{+17.4}_{-4.9}\times10^9$ \Msun, implying a stellar halo mass fraction of $M_{*,halo}/M_{*,tot} = 15^{+33}_{-9}$\%, on the high end for spiral galaxies, but with significant uncertainty. Finally, we find that M96 appears offset from the stellar halo mass--metallicity relationship for spirals, with a halo that is distinctly metal-poor for its halo mass. While a variety of systematic effects could have conspired to drive M96 off this relationship, if confirmed our results may argue for a markedly different accretion history for M96 compared to other spirals in the nearby universe.

\end{abstract}

\keywords{Galaxy Stellar Halos --- Stellar Populations --- Spiral Galaxies}

\section{Introduction}

The stellar halos of spiral galaxies contain a wealth of information
about the accretion history of their host galaxies. While some fraction
of a galaxy's stellar halo is built up in-situ, halos continue to grow
through the continual accretion and stripping of satellite galaxies over
time \citep[\eg][]{bullock05, font06, delucia08, cooper10, deason16}.
These accretion events deposit stars with a range of metallicity and age
into the halo \citep{belokurov06, ibata14, martin14, martin22,
bonaca25}, leading to significant spatial and kinematic substructure in
the stellar halo which slowly mixes away over time. These accretion
signatures are often most readily seen in the outer halos of galaxies
($\gtrsim$ 20--30 kpc) where the dynamical times are long and the
contrast with the light of the host galaxy is greater.

While the stellar halos of the Milky Way and Andromeda have been studied
extensively over time
\citep[\eg][]{ibata01a,ibata07,carollo07,mcconnachie09,carollo10,gilbert12,
gilbert14, ibata14, crnojevic17,mcconnachie18, conroy19, ogami25} , it
is only in recent years that observations have begun to probe the
stellar halos around more distant spiral galaxies. Deep, low surface
brightness imaging has revealed a myriad of tidal streams embedded in
the halos of massive spirals \citep[\eg][]{malin97, shang98, martinez10,
chonis11, merritt16, martinez23} tracing recent accretion events that
continue to build their stellar halos. Meanwhile imaging of the discrete
stellar populations in other galaxy halos from the Hubble Space
Telescope (and, for nearby galaxies, ground based observatories) have
begun to probe the metallicities (and in some cases, ages) of stars in
the halos of large spiral galaxies beyond the Local
Group\citep{mouhcine05a,mouhcine05b,rejkuba09,barker09,durrell10,
bailin11,rs11,monachesi16,cohen20,smercina20,gozman23}.

These studies have found a wide diversity in halo properties, with halo
stellar mass fractions ($M_{halo,*}/M_{gal,*}$) ranging from $<$ 1\% for
M101 \citep{merritt16,jang20, gilhuly22} to 2-3\% for the Milky Way and
M81 \citep{harmsen17,deason19,smercina20,mb20}, to over $\approx$ 10\%
for other luminous spirals \citep{harmsen17,bell17,smercina22}. Halo
stellar populations also show a wide range of metallicity, ranging from
$[M/H] = -2.0\ {\rm to} -0.1$
\citep{mouhcine05a,kalirai06,mouhcine07,durrell10, monachesi16,
harmsen17, cohen20,kang20,smercina22, gilbert22, gozman23}, and appear
generally old ($>$ 8 Gyr) but with some evidence for intermediate age
populations as well \citep{greggio14, rejkuba22, harmsen23}.

Taken together, these studies have revealed a striking and rather tight
correlation between the stellar halo mass and halo metallicity in spiral
galaxies \citep{harmsen17, bell17, smercina22}. Galaxies with anemic
stellar halos have relatively low halo metallicities, while those with
more luminous stellar halos show much higher halo metallicities. These
results suggest that the properties of galaxy stellar halos are most
strongly sensitive to the mass of the most massive merger(s) in the
galaxy's past accretion history, a behavior also seen in galaxy
formation simulations \citep{bullock05,cooper10,deason16, dsouza18a,
elias18, monachesi19,fattahi20}. In this context, the Milky Way and
Andromeda appear to represent opposite extremes on the scale of
accretion-driven halo properties. Andromeda's stellar halo is massive
\citep[$M_*\sim 10^{10}M_{\odot}$;][]{ibata14} and high in stellar
metallicity, and contains a wealth of complex substructure signifying an
active accretion history
\citep{ibata01a,ibata04,ibata07,tanaka10,gilbert12,gilbert14,ibata14,mcconnachie18,dsouza18b,dey23}.
In contrast, the halo of the Milky Way is lower in stellar mass
\citep[$\sim 10^{9}M_{\odot}$;][]{deason19,mb20} and metallicity, and
while it contains numerous accretion signatures as well \citep[\eg][and
references
within]{belokurov06,belokurov18,helmi18,naidu20,helmi20,db24}, they tend
to be lower in luminosity and than those found in Andromeda. A variety
of arguments suggest that the Milky Way has had a much more quiescent
accretion history than Andromeda, possibly leading to the stark
differences in their stellar halos \citep[\eg][]{deason13,
pillepich14,dsouza18a,sotillo22,wempe25}.

If accretion dominates the properties of stellar halos around galaxies,
differences in local environment may lead to evolution of, or scatter
in, the stellar halo mass-metallicity relationship, as ongoing
interactions and accretion continue to build galaxy halos. Within this
context, here we study stellar populations in the outer halo of the
spiral galaxy M96 (NGC~3368) at a distance of 11 Mpc \citep{graham97,
lee16}. M96 is the most luminous spiral in the Leo I galaxy group, a
group with a complex dynamical history. The group's massive elliptical
M105 (NGC~3379) is surrounded by the extended Leo \hi\ Ring
\citep{schneider85, schneider86}, possibly formed during an interaction
with M96 \citep{micheldansac10}. M96 itself shows a distorted outer \hi\
disk and is linked to the Leo Ring by a bridge of \hi\ gas
\citep{oosterloo10}. A number of \hi\ clouds, possibly tidal in origin,
are found between M96 and M95 \citep{taylor22}, the other large spiral
in the group, and the group is also home to the gas-rich ultradiffuse
galaxy BST1047+1156 \citep{mihos18bst,mihos24}, likely spawned from the
tidal debris connecting M96 and the Leo Ring. M96 thus provides an
interesting opportunity to study the stellar halo of a galaxy in a
dynamically active group.

In this study we use deep {\sl Hubble Space Telescope} (HST) imaging to
image two fields in M96's outer halo, at a projected distance of
$\approx$ 50 kpc from the center of the galaxy. Our imaging extends
nearly two magnitudes below the tip of the red giant branch (TRGB), and
we have a clear detection of red giant branch (RGB) stars in both
fields. We use the properties of this RGB population to derive the
galaxy's stellar halo mass and metallicity, and show that M96 appears an
outlier in the mass-metallicity relationship derived for other spiral
galaxy halos, having a distinctly metal-poor halo despite its
relatively large halo mass. 

\section{Observational Data}

\begin{figure*}[]
\centerline{\includegraphics[width=7.0truein]{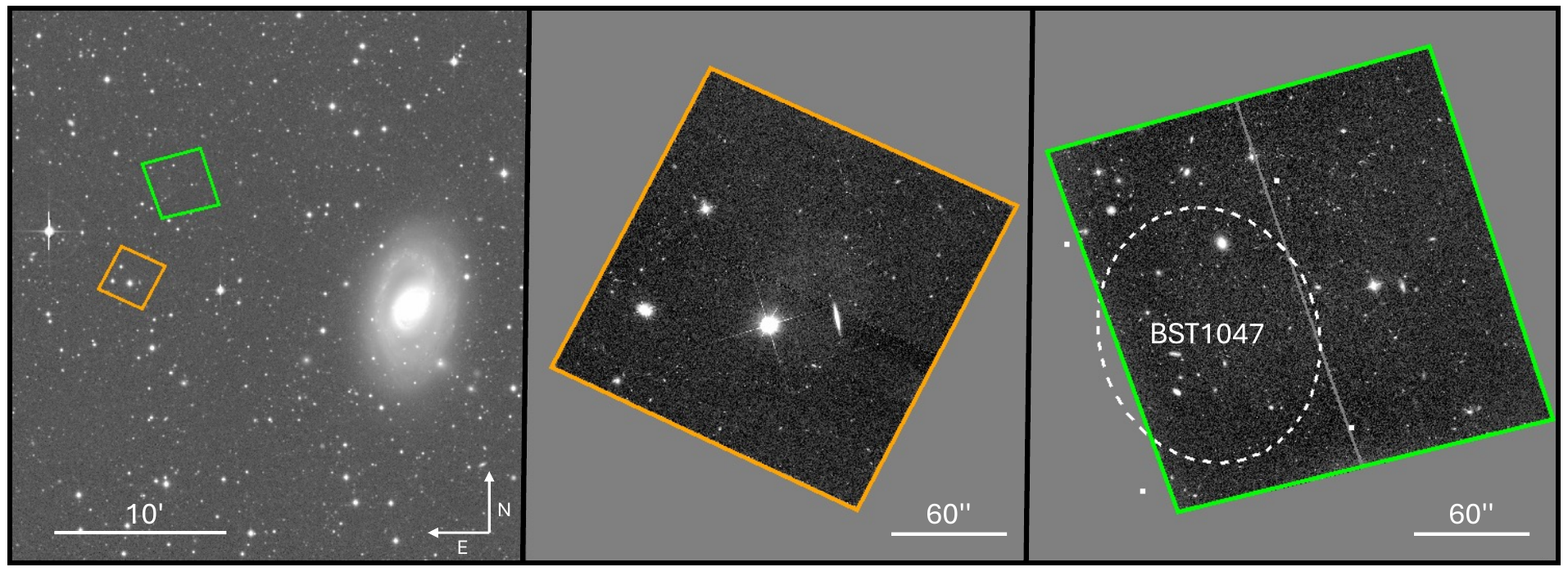}}
\caption{Placement of the WFC3 and ACS3 imaging fields. The left panel
shows the ACS (green) and WFC3 (orange) footprints overlaid on the
B-band imaging of M96 from \citet{watkins14}. The WFC3 and ACS3 images
are shown in the center and right panel, respectively. The dotted
ellipse in the ACS image shows the location of the ultradiffuse galaxy
BST1047+1156 \citep{mihos24}. The ACS field lies at a projected distance
of 15.2\arcmin\ (48.6 kpc) from the center of M96, while the WFC3 field
lies at a distance of 16.3\arcmin\ (52.2 kpc). In all images north is up
and east is to the left.
}
\label{imaging}
\end{figure*}

We imaged fields in the halo of M96 as part of our Hubble program
GO-16762 to study the resolved stellar populations in the ultradiffuse
galaxy BST1047+1156 \citep[hereafter BST1047][]{mihos18bst,mihos24}. In
taking that data, we placed BST1047 on the eastern half of the ACS field
(leaving the western half empty to sample a local background), while
also using the WFC3 camera in parallel to image an adjacent field. These
two fields, shown in Figure~\ref{imaging}, lie at a projected distance
of 15.2\arcmin\ (ACS) and 16.3\arcmin\ (WFC3) from the center of M96,
sampling the stellar populations in M96's halo at a projected physical
distance of $\approx$ 50 kpc \citep[using the adopted 11.0 Mpc distance
to the Leo I Group;][]{graham97,lee16}.

Both fields were imaged in the F606W and F814W filters over the course
of 15 orbits. For the ACS field, the total exposure times were 16458s
(F606W) and 16590s (F814W), while the exposure times for the WFC3 field
were 17500s (F606W) and 18750s (F814W). Two exposures were taken in each
orbit, and within each visit we utilized a small ($\sim$0.2\arcsec)
dither between exposures to aid in sub-pixel sampling of the imaging.
Larger dithers ($\sim$ 1\arcsec) were made between visits to avoid
artifacts on the detectors and to facilitate efficient cosmic ray
removal. We used the {\tt drizzlepac} package to construct deep,
drizzled images of each field. Before drizzling we used {\tt tweakreg}
to precisely align the world coordinate systems of the different images
using point sources on the individual {\tt flc} images, then used {\tt
astrodrizzle} to create stacked F606W and F814W images of each field;
these stacked images are shown in Figure~\ref{imaging}.

Rather than doing photometry on the drizzled stacks, we use the stacks
as reference images to identify point sources in each field, then use
the software package DOLPHOT \citep[an updated version of
HSTPhot;][]{dolphin00} to perform point-source photometry of these point
sources on the individual CTE-corrected {\tt flc} images. Here we give a
brief summary of the DOLPHOT photometric analysis; complete details of
the process can be found in \citet{mihos24} for the ACS imaging, and our
analysis of the WFC3 imaging follows that process closely.

We begin the data reduction by preprocessing the images using the
DOLPHOT {\tt acsmask} and {\tt wfc3mask} tasks to apply bad-pixel masks
and pixel-area masks to each image, then use {\tt splitgroups} and {\tt
calcsky} to break each image up into the individual instrumental chips
and calculate a preliminary background sky map for each chip image. In
each field, we then used DOLPHOT to perform object detection and
photometry simultaneously on the individual {\tt flc} images, using the
deep F814W drizzled image stack as the reference. The instrumental
magnitudes were then converted to the VEGAMAG photometric system using
updated zeropoints from the ACS zeropoint calculator
(https://acszeropoints.stsci.edu/). We also use DOLPHOT to inject and
measure 100,000 artificial stars (over the magnitude range $22 < {\rm
F606W} < 30$ and color range $-0.5 < {\rm F606W-F814W} < 2.0$) in each
of the ACS and WFC3 images in order to characterize the limiting depth
of the images as well as any photometric bias in the derived magnitudes.
All photometry reported here uses the VEGAMAG system, corrected for
foreground Galactic extinction using $A_{\rm F606W}=0.062, A_{\rm
F814W}=0.038$ \citep{schlafly11}.

To achieve the cleanest, most accurate photometric catalogs of point
sources in our deep imaging, we apply a variety of cuts to the DOLPHOT
photometric catalogs. First, we use Astropy's {\tt
photutils/segmentation} task \citep{photutils} to identify and mask
large objects in the field (bright foreground stars and background
galaxies), then reject sources found within this mask. We also mask
regions near the chipgap on the ACS and WFC3 detector where the
signal-to-noise is reduced by dithering.

We next use the photometric information provided by DOLPHOT to further
clean the photometry. Good sources need to be DOLPHOT TYPE=1 (``good
star'') with minimal crowding (CROWD$<$0.25) and have signal-to-noise
S/N$>$3.5 and goodness of fit CHI$<$2.4 in both the F606W and F814W
filters. We also apply a magnitude-dependent cut on the SHARP parameter
to reduce the severe contamination due to background sources at
magnitudes fainter than F814W$\approx$26. Following our earlier studies
of resolved stellar populations in Leo, Virgo, and the M101 group
\citep{mihos18m101,mihos22, mihos24}, we use a functional form for the
cut of $| {\rm SHARP} | < {\rm SHARP0} + 0.3e^{(m-m_{\rm crit})}$. Due
to the differences in pixel scale and depth between the imaging in the
different cameras and filters, the values of the parameters can be a
function of both camera and filter. For ACS we use the same parameters
used in our earlier analysis of the data \citep{mihos24}: ${\rm
SHARP0}=0.04, m_{\rm crit,F606W}=29.5, m_{\rm crit,F814W}=28.7$. Based
on our artificial star tests, we found that point sources in the WFC3
imaging appeared to show slightly non-zero SHARP values, so we adjusted
the WFC3 SHARP0 cut parameter to ${\rm SHARP0}=0.06$, broadening the
SHARP selection slightly, but otherwise left the $m_{\rm crit}$
parameters unchanged.

We used the artificial star tests to characterize the limiting depth
(defined as 50\% recovery rate) and systematic biases in the DOLPHOT
photometry. We process the artificial stars using the same selection
cuts used for the real data, and extract the completeness fraction, mean
error, and systematic shift in magnitude and color as a function of both
F814W magnitude and F606W$-$F814W color. Because of our joint selection
in F606W and F814W magnitude, completeness is a function of both
magnitude and color. For the ACS imaging, the 50\% completeness limit is
F814W=28.2 at blue colors (at F606W$-$F814W=0.0), and rises to
F814W=27.8 in the red (at F606W$-$F814W=1.0) For the WFC3 imaging, the
50\% completeness limits are F814W=28.2 (blue) and 28.0 (red). At
brighter magnitudes (F814W$\lesssim$27) we see no systematic shift in
derived magnitudes or colors, but at fainter magnitudes DOLPHOT
photometry is measured systematically too faint, by about 0.1 mag in ACS
and by as much as 0.15--0.2 mags in WFC3. These shifts are consistent
with our previous analysis of deep HST imaging \citep{mihos18m101,
mihos22, mihos24}, and the systematics are similar in both F606W and
F814W, such that the derived F606W$-$F814W colors show smaller
systematic offsets. In particular, in the region defining red giant
stars in the halo of M96 (see Section 3), the shifts are $\approx$ 0.05
mag in F814W magnitude and 0.03 or less in F606W$-$F814W color.

Finally, to avoid contamination from stellar populations in the
ultradiffuse dwarf galaxy BST1047 \citep{mihos18bst, mihos24}, we need
to mask that galaxy from our ACS photometry. We construct a large
elliptical mask on the eastern side of the ACS field (see
Figure~\ref{imaging}) that spans the entirety of BST1047. The mask
covers 36\% of the ACS field, reducing its effective area to 26,400
arcsec$^2$, comparable to that of the full WFC3 field (27,670
arcsec$^2$).

\section{Analysis}

\begin{figure*}[]
\centerline{\includegraphics[width=7.5truein]{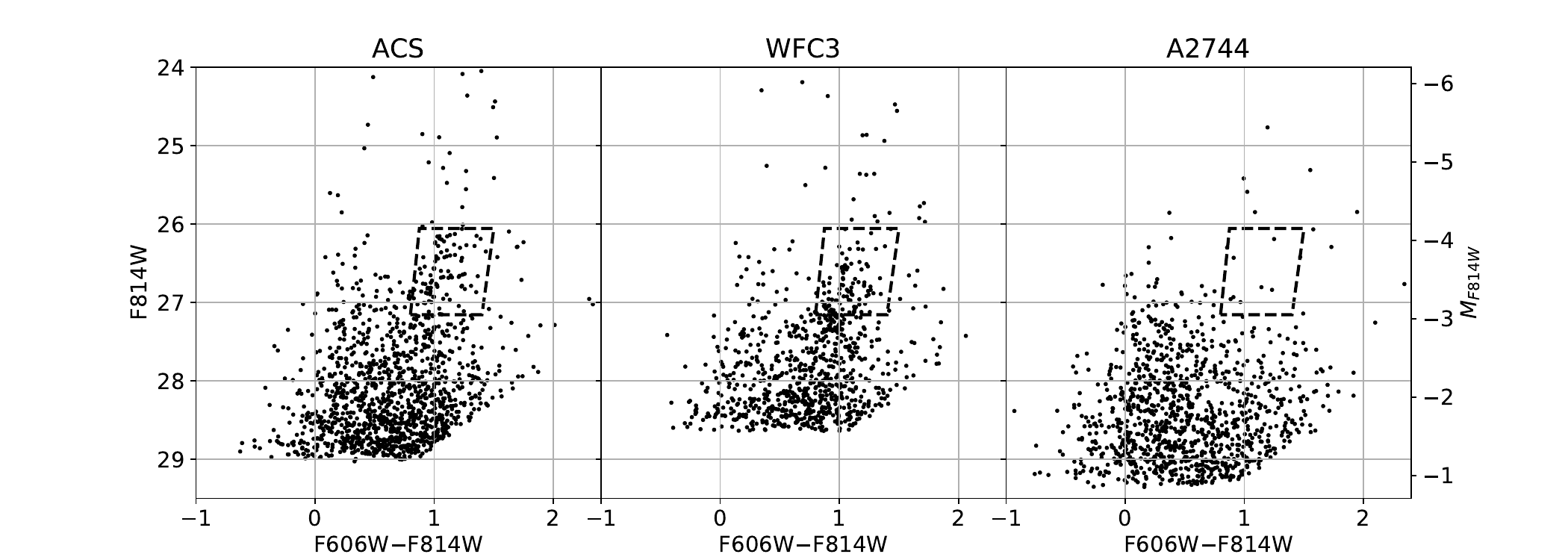}}
\caption{Color-magnitude diagrams for our ACS (left) and WFC3 (center) fields,
along with a similarly extracted CMD for the Abell 2744 Flanking Field (right).
The latter dataset has been randomly subsampled down by a factor of 0.6
to account for its larger area (see text). The dashed box in each CMD shows
the region expected to be populated by metal-poor red giant stars at the 
distance of M96.
}
\label{cmds}
\end{figure*}

We show the color-magnitude diagrams for point sources in the ACS and
WFC3 fields in Figure~\ref{cmds}. For comparison, Figure~\ref{cmds} also
shows a comparably deep color-magnitude diagram for point sources
derived from imaging of the Abell 2744 Flanking Field \citep[hereafter
A2744FF,][]{lotz17}. As described in \citet{mihos18m101}, the A2744
imaging provides a good control field to assess the expected background
counts as a function of color and magnitude in our deep studies of
resolved stellar populations in nearby galaxies such as M96. We analyse
that data set using the same techniques described above: using DOLPHOT
to extract point source photometry, then applying the same photometric
cuts as the M96 halo fields analysed here \citep[for more details
see][]{mihos18m101}. We note, however, two differences between the
A2744FF imaging and our M96 halo fields: that the A2744FF imaging has
$\sim$ 60\% larger area than the WFC3 or (masked) ACS fields, and sits
at higher Galactic latitude ($b_{A2744}=-81\degr$, compared to
$b_{M96}=+57\degr$). To account for the difference in field area, we
randomly select 60\% of the sources in the Abell2744FF dataset to plot
so that the density of sources shown on the CMD is comparable to that in
our M96 ACS and WFC3 fields.

The dashed box in Figure~\ref{cmds} shows the expected location of
metal-poor ([M/H]$\lesssim-0.7$) red giant stars at the distance of M96.
In that region of the CMD we see a clear excess of sources in both the
ACS and WFC3 fields compared to the A2744FF data. These sources have a
spatial density of 13.0 arcmin$^{-2}$ in the ACS field and 12.7
arcmin$^{-2}$ in the WFC3 field, compared to only 1.3 arcmin$^{-2}$ in
the A2744FF data. We expect little contamination from Milky Way stars;
using the Besan\c{c}on Galactic structure model \citep{robin03}, the
density of Milky Way halo stars in this region of the CMD is only 0.25
arcmin$^{-2}$. Thus we have a clear detection of M96 halo stars with
similar spatial density in both the ACS and WFC3 fields.

\begin{figure}[]
\centerline{\includegraphics[width=3.5truein]{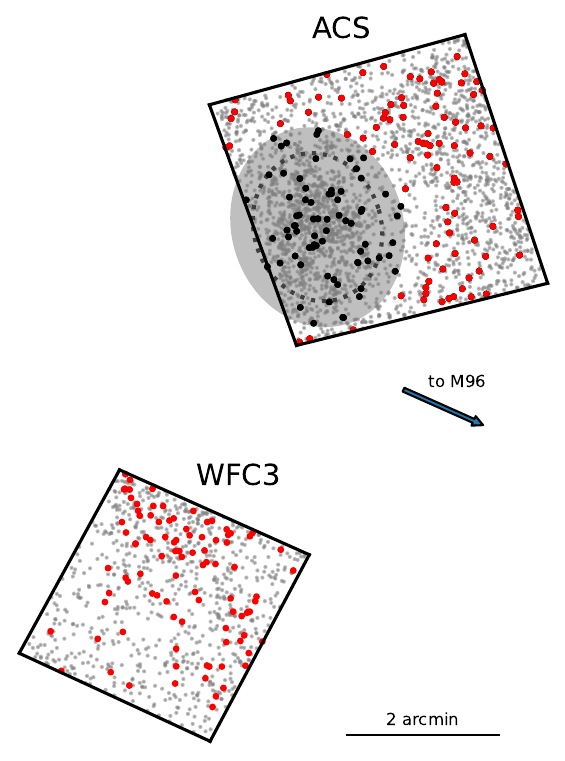}}
\caption{Spatial distribution of the stellar populations in the ACS and WFC3
fields. The light grey points show all detected point sources, while red dots
show red giant stars selected from the dashed RGB boxes in Figure~\ref{cmds}.
In the ACS field, the dotted oval shows the region containing BST1047, and the 
shaded oval shows the (slightly expanded) mask to exclude sources in this region.
However, for clarity in showing the overall distribution of RGB stars in the field, the
black dots show RGB stars within the masked region not included in our analysis.
In this figure, north is up and east is to the left.
}
\label{spatplot}
\end{figure}

Contamination from stellar populations associated with BST1047 also
appears unlikely. As shown in \citet{mihos24}, BST1047 shows prominent
red and blue helium burning sequences that trace a weak burst of star
formation in that object approximately 100--200 Myr ago. However, we see
no evidence for this population in our M96 halo photometry; the density
of sources in the CMD at bluer colors, where blue helium burning stars
would lie ($0.0 \leq {\rm F606W-F814W} \leq 0.6$ and $26 \leq {\rm
F814W} \leq 27$) is comparable in both the ACS and WFC3 fields (3.5 and
2.4 arcmin$^{-2}$, respectively), and similar to that observed in the
A2744FF CMD (3.0 arcmin$^{-2}$). Furthermore, as seen in
Figure~\ref{spatplot} the spatial distribution of sources in the RGB box
shows no clustering around BST1047 in the ACS field, arguing that these
sources are not red helium burning stars in BST1047. We also note that
subdividing the RGB population by color or luminosity does not change
the overall spatial distributions of the selected sources shown in
Figure~\ref{spatplot}. Finally, contamination by old stellar populations
from BST1047 is also not a concern; as shown in \citet{mihos24}, BST1047
has no old stellar population, and the distribution of RGB stars across
the full (unmasked) ACS field appears quite uniform \citep[as shown also
in Figure~9 of][]{mihos24} rather than being clustered near BST1047.

Another complication to our study of M96 halo populations is the group
environment, and the possibility of contamination by intragroup stars
stripped from other galaxies in the group. The presence of stellar
streams in the intragroup light could lead to field-to-field variations
in the inferred density of the stellar halo and in the ages or
metallicities of the stellar populations. However, deep imaging of the
Leo Group shows little evidence for intragroup light around M96 down to
a limit of $\mu_V = 29.5$ mag arcsec$^{-2}$ \citep{watkins14}. We also
see no difference in the spatial density or F606W$-$F814W colors of the
RGB stars in our two fields that probe the M96 halo. We do, however, see
a modest north-south gradient in the density of RGB stars across the
WFC3 field (shown in Figure~\ref{spatplot}). This gradient is
modest but appears statistically significant. We can compare the spatial
distribution of the data to that for a uniform distribution using the
$\sigma$/total metric of \citet{bell08}, which calculates the variation
in counts across spatial bins compared to the uniform distribution,
while properly factoring in the Poisson sampling statistics of the data.
Breaking the WFC3 field into quadrants to measure the statistic, we find
$\sigma$/total = 0.52$\pm$0.03 (compared to values of $\sigma$/total =
0.2$\pm$0.1 for a purely random distribution), arguing that the gradient
is real and may be tracing substructure around M96; however, this
gradient is weak, and additional deep fields around M96 would be
necessary to test the field-to-field variation in RGB surface density
more robustly. 

Comparing the RGB stars in the ACS and WFC3 fields, both show a broad
color spread indicative of a range of metallicities in the stellar
populations. Stars in RGB selection box for the ACS field show a
median color of F606W$-$F814W$=1.05\pm0.02$ (where the errorbar comes
from bootstrap resampling of the data), while those in the WFC3 field
show a median color of F606W$-$F814W$=1.07\pm 0.02$. The difference in
median color is very small, and consistent with the uncertainty in the
relative photometric zeropoint differences between the WFC3 and ACS
cameras \citep{deustua18}. Both fields also show very similar color
spreads. RGB stars in the ACS field show a F606W$-$F814W interquartile
spread of 0.20 mag, compared to a spread of 0.19 mag in the WFC3 field.
This color spread is significantly larger than the photometric color
uncertainty at these magnitudes ($\approx$ 0.07 and 0.10 for ACS and
WFC3, respectively), and thus reflects a true underlying metallicity
spread in the stellar populations. The similarity in median color and
color spread between the two fields argues that both fields are tracing
similar stellar populations.

\begin{figure*}[]
\centerline{\includegraphics[width=6.5truein]{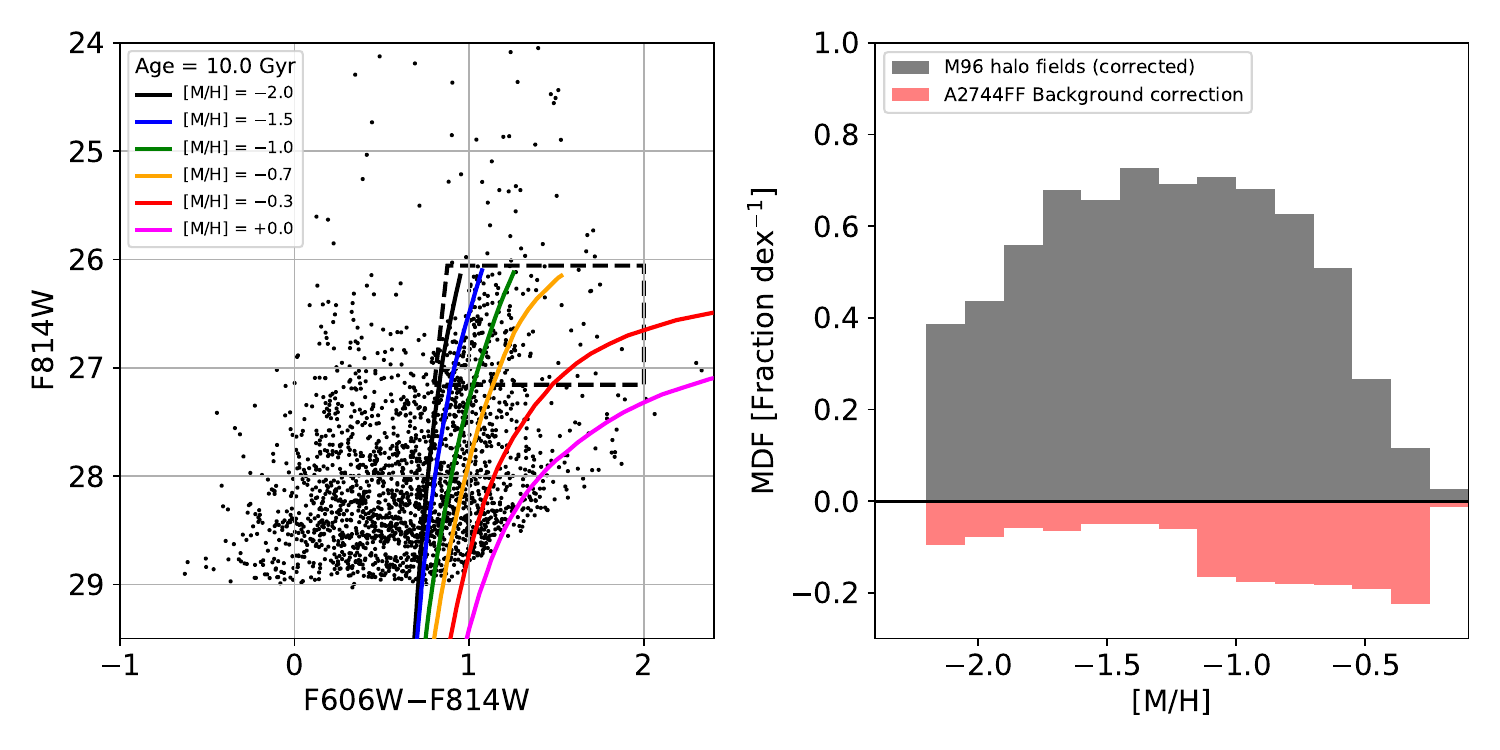}}
\caption{Left: The color-magnitude diagram for the combined ACS and WFC3
fields, overlaid with 10 Gyr old Parsec 1.2S isochrones \citep{bressan12, marigo17} of varying
metallicity. Right: the inferred M96 halo metallicity distribution
function (MDF) for the RGB population selected from stars in the dashed
box of the CMD. The halo MDF has been corrected for background
contamination based on an analysis of similarly-selected sources in the
Abell2744 Flanking Field photometry, shown in red. See text for details.
}
\label{MDFs}
\end{figure*}

With no discernable difference in the color distribution of the two
fields, we combine the ACS and WFC3 photometry and replot them together
in Figure~\ref{MDFs}a, overlaying Parsec 1.2S isochrones
\citep{bressan12, marigo17} for 10 Gyr old stellar populations spanning
a range of metallicity ${\rm [M/H]}=-2.0\ {\rm to}\ 0.0$. Although we
are largely insensitive to stars of solar metallicity, we see little
evidence for stars of metallicity higher than ${\rm [M/H]}=-0.5$. There
are very few stars in the CMD redward of F606W$-$F814W$=1.5$, and what
few stars are there are consistent with the expectations for
contaminants from the A2744FF data (Figure~\ref{cmds}).

We use these isochrones to estimate the metallicity distribution
function (MDF) for the M96 halo populations. We first select all point
sources in the RGB selection box shown in Figure~\ref{MDFs}a, which has
been expanded redwards compared to that shown in Figure~\ref{cmds} to
account for any metal-rich population. Then, using a grid of finely
interpolated Parsec isochrones of age 10 Gyr and spanning the range
${\rm [M/H]}=-2.2\ {\rm to}\ -0.2$, for each observed RGB star we locate
the closest model isochrone and assign the star the metallicity of that
isochrone. We account for photometric uncertainty by repeating this
process 100 times, each time randomly scattering the isochrone
points in color and magnitude by adding a Gaussian random error based on
the photometric uncertainty model. To correct for contamination, we run
an identical analysis on sources in the A2744FF field and subtract the
resulting MDF from that derived from the combined ACS and WFC3
photometry. We show the derived background-corrected metallicity
distribution function (MDF) for M96's halo in Figure~\ref{MDFs}b, along
with the background correction inferred from the A2744FF data.

The M96 halo MDF shows a broad distribution in metallicity spanning
the range $-2.2 \leq {\rm [M/H]} \leq -0.3$. The sharp cutoff at [M/H]
$= -2.2$ is formally due to the lack of isochrones at lower
metallicities, but we also note that at these extremely low inferred
metallicities the true counts have greater uncertainty due to increasing
contamination from unresolved background galaxies on the blue side of
the RGB selection box. The amount of contamination in this region of the
CMD is highly sensitive to the exact details of how sources are selected
using the DOLPHOT SHARP cuts described in \S2, and is reflected in the
increased contamination correction at very low metallicity inferred from
the A2744FF field as well (the red histogram in Figure~\ref{MDFs}b).
Thus we do not place much significance to this lowest metallicity bin in
the derived MDF. If we exclude this bin from the analysis of the MDF, we
find a median metallicity for the M96 halo of [M/H] $=-1.36$, with an
interquartile spread of 0.75 dex. This metallicity is similar to that
found in the outer halos of other spiral galaxies \citep{harmsen17,
conroy19, smercina20}, and we discuss this more fully in Section~4
below.

We estimate the uncertainty in the inferred median metallicity in
two ways. First, to estimate the random errors we run a bootstrap
analysis, calculating the median metallicity on 100 subsamples of the
RGB stars randomly drawn (with replacement) from the full RGB sample; we
find a random error in the median of $\approx\pm0.05$ dex. Systematic
uncertainties are harder to assess. As one example, we recalculate the
metallicities using isochrones that vary in age from 6 Gyr to 12 Gyr and
find that the derived median metallicity varies along this sequence from
[M/H]=$-1.19$ to $-1.40$. Alternatively, we can bypass the model
isochrones and instead use the color-metallicity calibration of
\citet{streich14} (based on the metallicities of Galactic globular
custers) to convert the median F606W$-$F814W color of the RGB stars to
metallicity. Using the \citet{streich14} calibration at an absolute
F814W magnitude of $M_{\rm F814W}=-3.5$ (F814W=26.7), we get
[M/H]=$-1.42$, while using the $M_{\rm F814W}=-3.0$ calibration we get
[M/H]=$-1.20$. These calculations suggest a {\it minimum} systematic
uncertainty of $\pm0.1$ dex, and likely even higher \citep[see, \eg the
discussion in][]{streich14}.

\section{Discussion}

Our imaging of the outer halo of M96 has revealed a sparse population of
old red giant branch stars with a broad range of metallicity at a
galactocentric distance of 50 kpc. While the ACS and WFC3 fields do not
sample different halo radii, they are separated by about 19 kpc and can
probe the uniformity of the halo stellar populations. We find little
variation in their properties, however. Both fields have similar stellar
densities (11.7 arcsec$^{-2}$ vs 11.4 arcsec$^{-2}$ for ACS and WFC3,
respectively, after background correction), median RGB color (1.05 vs
1.07), and RGB color quartile spread (0.20 vs 0.19), with all values
agreeing to well within the uncertainties. There is some evidence for a
weak north-south gradient in RGB counts in the WFC3 field (see
Figure~\ref{spatplot}), but aside from that possibility, we find no
obvious evidence for substructure in the stellar populations.

We can use the counts to estimate the equivalent projected surface
brightness of the halo. Using the PARSEC isochrones for an old (10 Gyr)
metal poor ([M/H] $=-1.3$) stellar population and scaling in mass to
match the RGB number counts seen in our imaging fields, we derive an
equivalent surface brightness of $\mu_V \approx 31.7$ mag arcsec$^{-2}$
for these fields. This is consistent with the non-detection of diffuse
light in this region by \citet{watkins14}, since the equivalent surface
brightness derived from our HST imaging is significantly fainter than
the \citet{watkins14} surface brightness limit of $\mu_V \approx 29.5$
mag arcsec$^{-2}$.

To test how well the M96 stellar halo follows the stellar halo mass --
metallicity relationship for spiral galaxies we need an estimate of M96's
stellar halo mass. Previous studies \citep[\eg][]{harmsen17, gozman23}
have typically measured this by using data taken at different radii to
derive the halo density profile, then integrating that projected profile
from 10--40 kpc (to avoid contamination from the bright inner regions of
the galaxy). However, without fields that span a range of radii we
cannot characterize the halo density profile, so instead we adopt a
power law profile for the projected stellar density, given by $\Sigma(R)
= \Sigma(R_0) (R/R_0)^\alpha$. Parametrized in this way, spiral galaxies
show an observed variety of halo profile slopes spanning the range
$\alpha = -2$ to $-4$
\citep{deason11,gilbert12,ibata14,harmsen17,mb20,smercina20,gozman23}.
The integrated mass will also depend on the flattening of the halo,
but again, with only two fields projected along the minor axis of M96
we cannot measure the halo flattening. Instead, we adopt a projected
axis ratio of $b/a=0.6$, similar to the observed axis ratios for stellar
halos around nearby spirals \citet{harmsen17}. Normalizing the profile
to our measured projected density at 50 kpc ($\Sigma = 11.65$ RGB stars
arcmin$^{-2}$) we then integrate the profile over the radial range
corresponding to 10--40 kpc in projection to derive a halo mass
over that radial range of $M_{h,10-40} = 2.6^{+5.8}_{-1.7} \times
10^{9}$ \Msun, where the errorbar shows the effect of varying the
profile slope over the range $\alpha = -2$ to $-4$. We also note that
simulations suggest some correlation between halo density profile slope
and accretion history \citep{monachesi19}, so the range in slope we
consider can also be considered as spanning a range of accretion
scenarios. Finally, we need to convert the $M_{h,10-40}$ halo mass
to a total stellar halo mass. We follow the conversion from
\citet{harmsen17} who used the properties of simulated stellar halos in
\citet{bullock05} to show the the total stellar halo mass is roughly a
factor of three larger than $M_{h,10-40}$. Using this scaling factor, we
ultimately derive a total stellar halo mass for M96 of
$M_{h,tot}=7.8^{+17.4}_{-4.9}\times 10^{9}$ \Msun.

\begin{figure}[]
\centerline{\includegraphics[width=3.3truein]{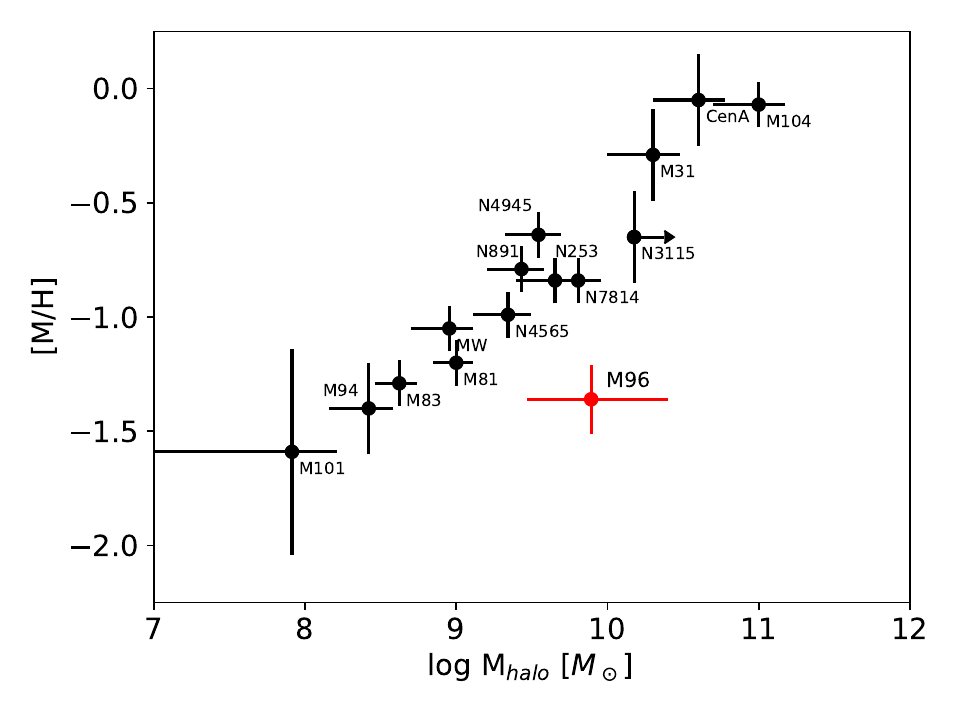}}
\caption{The halo mass -- halo metallicity relationship from \citet{gozman23},
with our measurement of M96 added. 
Note that the metallicity measurements for the \citet{gozman23} points are
measured at 30 kpc, while our fields in M96 are at a larger projected distance
of 50 kpc. 
}
\label{massmet}
\end{figure}

We use this estimate of halo mass to put M96 onto the halo mass -- halo
metallicity relationship shown in Figure~\ref{massmet}, using the data
from \citet{gozman23} which adds galaxies to the previous relationships
derived by \citet{harmsen17} and \citet{smercina22}. In our M96 datapoint, the metallicity
errorbar reflects a combination of the $\pm0.05$ dex random uncertainty
and the $\pm0.10$ minimum systematic uncertainty discussed in Section~3.
Our M96 measurement falls well below the mean relationship, by
$\approx 0.7$ dex in metallicity. Given the tight scatter in the \citet{gozman23}
dataset, the offset of M96 from the other galaxies is quite significant.

A variety of factors may contribute to this displacement of M96 from the
mean relationship between halo mass and metallicity. First, our
metallicity measurement is made at a projected radius of 50 kpc, while
the other galaxies in the plot are measured at 30 kpc. If there is an
appreciable metallicity gradient in M96's halo, this would push the M96
point upwards, more in line with the other galaxies. While spiral galaxy
halos show a range of metallicity gradients, to completely explain the
offset a substantial gradient would be needed ($\approx -0.035$
dex/kpc), much larger than is typically seen \citep[\eg][]{gilbert14, monachesi16,
gozman23}. Another concern is that our estimated halo mass may be too
high. With only two closely separated fields, our  mass estimate is
based on assumptions both about the halo density slope and its projected
flattening. A shallower slope and/or rounder halo would reduce the
inferred mass and again bring M96 more into alignment. Finally, we do
see some evidence for substructure in the projected counts, particularly
in the WFC3 field. If our fields happen to sample discrete tidal
streams, either in M96's halo or perhaps in the Leo Group intragroup
environment, this could again lead to an overestimate of M96's halo
mass, and potentially also bias the inferred metallicity in either
direction. Indeed, some combination of all three of these effects ---
metallicity gradient, halo density profile, and tidal substructure ---
could work together to move M96 back towards the mean relationship;
imaging of the stellar populations in additional fields around M96 would
be necessary to assess these variations.

However, if our estimates of the mass and metallicity of M96's halo are
confirmed, it would suggest that more scatter in the spiral halo
mass-metallicity relation exists than previously realized. As the
mass-metallicity relationship is often interpreted in the context of
accretion-driven growth of stellar halos \citep[see,
\eg][]{dsouza18a,monachesi19,smercina22}, this would in turn imply
differences in the assembly history of M96 compared to other spirals.
For example, M96 may have suffered a major merger particularly early in
its history, accreting a massive but relatively metal-poor satellite to
build its halo mass without a concurrent increase in metallicity.
Alternatively, over its history M96 may have accreted a particularly
large number of low mass, metal-poor satellites, growing a massive but
metal-poor stellar halo more gradually over time. The fact that stellar
halos reflecting such scenarios are not commonly seen in simulations
\citep{dsouza18a,monachesi19} would suggest that these are uncommon
evolutionary paths, consistent with M96's offset from the otherwise
relatively tight correlation between halo mass and metallicity. Finally,
while it may be tempting to consider M96's presence in the active Leo~I
group environment as shaping an alternative evolutionary path, we note
that many of the galaxies that define the mass-metallicity
relationship seen in Figure~\ref{massmet} are also found in group
environments, including M101, M81, and of course the Milky Way and
Andromeda.

While the metallicity distribution shown in Fig~\ref{MDFs} is broad, it
is not unreasonably so for the halo of a large spiral galaxy built up
through accretion. For example, in the Auriga galaxy formation
simulations \citep{grand17}, the outer regions of spiral galaxy halos
show a wide range of metallicities, with individual halos showing a
10\%--90\% metallicity spread of a dex or more
\citep{monachesi19,fattahi20}, compared to the 10\%--90\% metallicity
spread of 1.1 dex that we find in M96. Observed MDFs for stars in spiral
galaxy halos also tend to be broad
\citep[\eg][]{mouhcine05b,mouhcine07,rejkuba09,carollo10,ibata14}

Alternatively, we can directly compare the F606W$-$F814W color spread of
our M96 halo stars to the colors observed in other spiral galaxy halos
by the GHOSTS team \citep{monachesi16}. By comparing RGB star colors, we
alleviate systematic effects due to the transformation of colors to
metallicity. To calculate the color spread in the GHOSTS sample,
\citet{monachesi16} define a color quantity known as ``Q-index color'',
which essentially rotates the CMD so that the metal-poor RGB sequence is
vertical in the rotated CMD. Done this way, the Q-index then minimizes
the the effect of the tilt of the RGB in the color magnitude diagram on
the derived color spread. Selecting RGB samples that vary in depth from
0.5--2 mag below the RGB tip, \citet{monachesi16} find Q-index spreads
(16\%--84\%) of 0.25--0.5 mags for the halos in the GHOSTS sample. For
our M96 measurement, we select stars up to one magnitude below the tip
($m_{\rm tip,F814W}=26.2$) and in the color range $1.0 \leq {\rm
F606W-F814W} \leq 1.4$, similar to the range used by
\citet{monachesi16}, and derive a Q-index spread of 0.27 mag for the M96
halo. This similarity in color spread between M96 and the GHOSTS halos
again argues that the spread of metallicities M96's halo is typical of spiral 
galaxies in general.

Finally, we can also use our estimate of M96's stellar halo mass
($M_{tot,halo} = 7.8^{+17.4}_{-4.9}\times10^9$\Msun) to estimate the
stellar halo mass fraction ($M_{tot,halo}/M_{tot,gal}$). If we adopting
a total stellar mass for M96 of $M_{*,tot} =5.3 \times 10^{10}$\Msun\
from the S4G survey \citep{sheth10,munoz13,querejeta15}, we derive
a stellar halo mass fraction of 15$^{+33}_{-9}$\%. Spiral galaxies show
a wide variation in halo mass fraction, ranging from $\leq 1\%$ for
galaxies such as M101 \citep{merritt16,jang20}, a few percent for the
Milky Way and M81 \citep[\eg][]{deason19,mb20,jang20, smercina20} to
10\% or more for galaxies such as M31, NGC~1084, and NGC~3368
\citep{harmsen17,bell17,smercina22}. Simulations of stellar halos formed
through accretion also show a wide range of mass fractions
\citep{cooper13,monachesi19}. Conversely, while M96 appears to fall in the
upper range of halo mass {\it fraction} for a large spiral galaxy, its
stellar halo itself is not anomalously massive. The total stellar halo
mass of the Milky Way (and M81) is smaller, $\sim 10^9M_\odot$
\citep{smercina20,mb20}, while M31 has a larger total stellar halo mass
of $M_{tot,halo,} \sim 10^{10}$\Msun \citep{ibata14, harmsen17}.  Although our
uncertainties are rather large, it thus appears that M96 has a stellar halo mass
that is  intermediate between these ``more typical''
spiral galaxies.

\section{Summary}

We have used deep HST ACS and WFC3 imaging in F606W/F814W to study the
resolved stellar populations in the outer halo of the spiral galaxy M96,
at a projected radius of 50 kpc from the galaxy's center. Our imaging
probes nearly two magnitudes below the expected tip of the RGB
luminosity function, and in both fields we have a clear detection of RGB
stars at a surface density (over the expected background) of $\approx$
11.65 arcmin$^{-2}$. Comparing the properties of the RGB stars in each
field, we find no significant differences in their surface density or
their mean F606W$-$F814W color or quartile color spread.

Combining the photometry from the two fields, we then use stellar
isochrones to derive the metallicity distribution function for the RGB
stars. We find a median metallicity of [M/H]=$-1.36$ with a
quartile spread of 0.75 dex. At this metallicity, the equivalent surface
brightness of M96's halo is $\mu_V \approx 31.7$ mag arcsec$^{-2}$, well
below the upper limits on the M96 halo surface brightness from deep
imaging of the Leo I Group by \citet{watkins14}. Assuming a power law
profile for the projected radial density of the halo, we derive a total
halo mass of $\approx 7.8^{+17.4}_{-4.9}\times 10^9$\Msun, approximately
15$^{+33}_{-9}$\% of the total stellar mass of the galaxy.

With our estimates of halo mass and metallicity in hand, we place M96 on
the stellar halo mass-metallicity relationship and find the galaxy falls
significantly below that relationship. Compared to other nearby spiral
galaxies, M96 appears to have a relatively metal-poor halo (by
$\sim$ 0.7 dex) for its halo mass. While the low metallicity we find for
M96's halo may be partly explained by the larger radius we probe
(compared to other galaxies that define that relationship), additional
systematic effects may be at work, including uncertainties in our
adopted halo mass profile and the possibility that we are sampling
underlying tidal streams in M96's halo or the Leo~I intragroup
environment. These systematic uncertainties aside, if our result is confirmed
it would imply that the accretion path that built M96's halo may be
systematically different from that inferred for other spirals in the
nearby universe. Possibilities to build up a massive metal-poor halo for
M96 include a very early merger with a massive but low metallicity
companion, or a more gradual accretion of a large number of low mass,
metal-poor satellites.

Furthermore, while it may be tempting to ascribe M96's outlier status to
its ongoing interactions within the dynamically active Leo I Group, it
is not obvious that this is correct. To illustrate this, an interesting
comparison can be made between M96, M81, and M101. All three of those
galaxies live in active group environments and have had recent
interactions within the past few hundred Myr --- M96, as traced by the
stellar populations in BST1047 \citep{mihos24}; M81 with M82 and NGC3077
\citep{yun94}; and M101 with NGC 5474 \citep{linden22}. While all three
are experiencing recent interactions, M101 and M81 both adhere to the
halo stellar mass - metallicity relationship seen in
Figure~\ref{massmet}, while M96 remains an outlier. This result
reinforces the idea that the halo properties of galaxies are likely more
sensitive to their integrated accretion history rather than their
immediate dynamical environment.

The main limitation to our study of the M96 stellar halo is the
sparseness of our imaging data. With only two fields surveyed, both at
similar galactocentric distance from M96, we cannot derive a radial
density profile nor can we test for the presence of substructure in the
halo. While both our fields have similar stellar populations and surface
densities, we do see some spatial variation in the form of a weak
density gradient in the WFC3 field. Whether or not this is signaling a
larger degree of underlying substructure is unclear, and further imaging
of the stellar populations in the field around M96 --- and throughout
the Leo I Group as a whole --- would shed new light on the role
interactions and accretion play in shaping the halos of spiral galaxies.

\begin{acknowledgments}

The authors would like to thank Christian Soto and Norman Grogin for
their help with planning and refining the HST observations. We also
thank Adam Smercina and Katya Gozman  for useful discussions and for
providing in digital form the data used in Figure~\ref{massmet}. We also
thank the anonymous referee for suggestions that significantly improved our
presentation of the results. This research is based on observations made
with the NASA/ESA Hubble Space Telescope for program \#GO-15258 and
obtained at the Space Telescope Science Institute (STScI). STScI is
operated by the Association of Universities for Research in Astronomy,
Inc., under NASA contract NAS5-26555. Support for this program was
provided by NASA through grants to J.C.M. and P.R.D. from STScI. A.E.W.
acknowledges support from the STFC (grant numbers ST/Y001257/1 and
ST/X001318/1).

\end{acknowledgments}

\facility{HST (ACS, WFC3)}. The {\sl Hubble Space Telescope} imaging data used
in this study can be accessed at the Mikulski Archive for Space
Telescopes (MAST) at the Space Telescope Science Institute via
\dataset[DOI: 10.17909/300a-ns91]{https://doi.org/10.17909/300a-ns91}.

\software{
astropy  \citep{astropy13, astropy18, astropy22}, 
DOLPHOT \citep{dolphin00},
numpy \citep{numpy},
matplotlib \citep{matplotlib},
photutils \citep{photutils}
scipy \citep{scipy},
}

\bibliographystyle{aasjournal}

\end{document}